\documentclass[12pt]{iopart}

\expandafter\let\csname equation*\endcsname=\relax 
\expandafter\let\csname endequation*\endcsname=\relax 
\usepackage{amsmath,amssymb,amsthm}

\usepackage{physics}
\usepackage{graphicx}
\usepackage{color}
\usepackage{hyperref}

\newcommand{\Ss}{_\textsc{s}}
\newcommand{\Sb}{_\textsc{b}}
\newcommand{\Ssb}{_\textsc{sb}}
\newcommand{\Ssh}{_\textsc{sh}}
\newcommand{\Ssc}{_\textsc{sc}}
\newcommand{\Sh}{_\textsc{h}}
\newcommand{\Sc}{_\textsc{c}}

\DeclareMathOperator{\sign}{sign}

\begin{document}

\title[An operator derivation of the Feynman-Vernon theory]{An operator derivation of the Feynman-Vernon theory, with applications to the generating function of bath energy changes and to an-harmonic baths}

\author{Erik Aurell $^1$, Ryochi Kawai$^2$,
Ketan Goyal$^2$\footnote{Present address:
Avigo Solutions, LLC
   1500 District Avenue, Burlington, MA 01803, USA}}
\address{$^1$ KTH -- Royal Institute of Technology,  AlbaNova University Center, SE-106 91~Stockholm, Sweden}
\address{$^2$ Department of Physics, University of Alabama at Birmingham, Birmingham, AL 35294, USA}
\ead{eaurell@kth.se}

\begin{abstract}
We present a derivation of the Feynman-Vernon approach
to open quantum systems in the language of super-operators.
We show that this gives a new 
and more direct derivation of the generating function
of energy changes in a bath, or baths.
As found previously, this generating function is given by a Feynman-Vernon-like
influence functional, with only time shifts in the kernels coupling the forward
and backward paths.
We further show that the new approach extends to an-harmonic and possible non-equilibrium baths,
provided that the interactions are bi-linear, and that the baths
do not interact between themselves.
Such baths are characterized by non-trivial cumulants.
Every non-zero cumulant of certain environment correlation
functions is thus a kernel in a higher-order term in the Feynman-Vernon
action. 
\end{abstract}

\maketitle

\section{Introduction}\label{sec:introduction}
When a quantum system interacts with an environment (open quantum system or OQS),  the state of the system is influenced by the environment in a fundamental manner.  Decoherence, for example, is  a consequence of quantum entanglement between the system and the environment~\cite{Schlosshauser2007}.  The understanding of effects on the system induced by the environments is essential to the quantum information technology~\cite{Wilde2017} and quantum thermodynamics~\cite{Binder2019}.  
A direct inclusion of the environment in a first principle investigation is usually not practically feasible. 
On the other hand, the structure and details of a large environment can only partially be reflected in the dynamics of the system.
In the super-operator approach going back to Nakajima and Zwanzig~\cite{Nakajima1958,Zwanzig1961} one 
starts from equations of motion (von Neumann-Liouville equations) of the total
density matrix of the system and the environment, and 
projects that to an effective dynamics for the system density matrix~\cite{Zwanzig2001,Breuer2002,Weiss2012}.
In the alternative approach of
Feynman and Vernon the influence of an environment on the system is expressed in terms 
of a influence functional~\cite{Feynman1963}.
The development of the reduced density matrix of the system
is then given by a double path integral,
where the influence functional couples the two paths.
If used exactly, both approaches agree and
specify completely positive dynamic maps describing the evolution of the system.
Once these are obtained, the system dynamics can be investigated without the knowledge of environment dynamics~\cite{Breuer2002,Weiss2012}. 

This paper is about the relation between the two approaches, and how important extensions obtained only recently are much more easily
derived in the super-operator approach. We will also show that the super-operator approach 
yields new higher-order corrections to Feynman-Vernon.
Before proceeding to the main argument, we note that 
even after one has found a closed form expression of a dynamical map or the influence functional, calculating the time-evolution of the system using them is technically challenging.  
Tractable methods using various approximations have been developed. Quantum master equations (Lindblad equations)~\cite{Lindblad1976,Alicki2010} based on the Born-Markovian approximation are the most popular, and can be derived in both approaches~\cite{Breuer2002,CaldeiraLeggett83a,Grabert88}.
Non-Markovian  methods are also developed~\cite{Breuer2016,deVega2017,GrifoniHanggi1998}. For the ``spin-boson'' problem of one two-state system (a qubit) interacting with a bosonic bath the non-interacting blip approximation (NIBA) was developed~\cite{Leggett87}, and shown to be equivalent to
relaxation after a polaron transform~\cite{Aslangul1986,Dekker1987}.  Many numerical algorithms have been developed
to treat the spin-boson problem with one or several bath including 
the hierarchical equation of motion (HEOM)\cite{Tanimura1989,Tanimura2014,Tanimura2015,Kato2015,Kato2016,deVega2017}, the quasi-adiabatic propagator path integral (QuAPI)\cite{Makri1998,Boudjada2014}, 
the multi-configuration time-dependent Hartree (MCTDH) approach~\cite{Velizhanin2008},
the Stochastic Liouvillian algorithm~\cite{StockburgerMak99}, and other Monte Carlo approaches~\cite{Saito2013}.

The thermodynamics of an OQS describes how a quantum system exchanges energy,
particles and other quantities with one or several reservoirs~\cite{Strasberg2017}.
While in Quantum Markov dynamics the energy interchanged with a reservoir, which we call \textit{heat},
can be expressed in terms of 
Lindblad operators acting on the system density matrix~\cite{Alicki1979}, in general that is not so.
Nevertheless, it has recently been shown by several groups that
the generating function of heat can be computed by the path integral technique
in a Feynman-Vernon-like approach~\cite{Carrega2015,AurellEichhorn,FunoQuan2018,Aurell2018}. In this formulation appears a 
new influence functional depending on the generating function
parameters. However, the terms in this new influence functional
are in fact the same as in Feynman-Vernon, with only a time shift in the argument in some of the kernels
as announced previously in~\cite{Aurell2018-Erratum}.
We will in this work derive the same result in the 
super-operator formalism where it emerges in a 
straight-forward manner without
cancellations in intermediate steps of the calculation.

We will also show that the super-operator approach 
can be extended beyond ideal Bose gas.
Cumulants of specific bath correlation functions (to be discussed
below) then enter as kernels in higher-order order
terms in the Feynman-Vernon action.
For instance, a non-zero third-order bath correlation function
gives the kernel of a third-order term in the Feynman-Vernon action.
This result appears more difficult to obtain in the path integral
formulation.

We present the equivalence of super-operator and Feynman-Vernon 
approach in a coherent manner through the generating function
of heat as an example, and the extension to an-harmonic baths.
The paper is organized as follows.
Section~\ref{sec:OQS} contains a brief summary of the path integral
and super-operator approaches to the development of
the system density matrix, and
Section~\ref{sec:generating-function}
extends the discussion to generating function of heat.
Section~\ref{sec:general-theory} contains a systematic and general
derivation of the super-operator approach, and
Section~\ref{sec:non-harmonic} the extension to an-harmonic baths.
Section~\ref{sec:discussion} sums up and discussed the results.
Appendix~\ref{sec:time-ordering-operator}
defines the super-operator time-ordering used in the
main body of the paper, and Appendix~\ref{sec:pair-correlation}
contains the details of the pair correlation functions in harmonic
baths.
Appendix~\ref{app:time-ordered-cumulants} shows the third- and fourth-order
kernels in high-order influence functional that result from non-zero
third-order and fourth-order cumulants in the bath.
Appendix~\ref{app:path-integrals-harmonic-baths} 
gives for completeness an outline of the path integral derivation
the result of which was
previously announced in ~\cite{Aurell2018-Erratum}.

\section{Theory of Open Quantum Systems: A Brief Summary}
\label{sec:OQS}
Consider a system S in contact with an environment B. Their Hamiltonians are denoted as $H\Ss$ and $H\Sb$, respectively and they are coupled through a interaction Hamiltonian $V\Ssb$. The whole system is assumed to be initially in a product state $\rho(t_i) = \rho\Ss(t_i) \otimes \rho\Sb(t_i)$ and evolves by a unitary transformation:
\begin{equation}\label{eq:full-evolution}
\rho(t) = U(t;t_i) \rho(t_i) U^\dagger(t;t_i)
\end{equation}
where $\rho$ is the density operator of the whole system and $U(t;t_i)=e^{-i H (t-t_i)}$ is a usual time-evolution operator with the total Hamiltonian $H=H\Ss+H\Sb+V\Ssb$.
Units of time are chosen such that $\hbar=1$.

When considered together with a time-constant interaction the assumption of an initial 
product state limits the analysis to weak system-bath interaction.
One way around that problem is to take the system-bath interaction time-dependent 
and small only initially,
as was allowed for in the original treatment by Feynman and Vernon~\cite{Feynman1963}.
Other approaches were discussed in~\cite{Grabert88},
and have some advantages for the analysis of the quantum state.
When considering thermodynamic quantities the question is 
more involved. Even in classical mesoscopic systems interacting strongly with an
environment the concept of heat is controversial, and has
been vigorously debated in the recent literature~\cite{Seifert2016,TalknerHanggi2016,Jarzynski2017,MillerAnders2017}.
As discussed by one of us~\cite{Aurell2017} different proposals for strong-coupling 
heat can be understood as different types of control and joint initial conditions
of bath and baths. 
On the quantum side, aspects of some of these issues were discussed some time ago 
in exactly solvable models~\cite{daCosta2000,Ingold2009}.
Here we will follow the option made available in~\cite{Feynman1963} and assume that the system-bath interaction 
vanishes at the beginning and the end of a process, but can be arbitrarily strong in between.
Heat can then be identified by the energy change in a bath.
Heat has been calculated in this scenario using Eq. \eqref{eq:inful-func-op} in \cite{Goyal2019}.

Returning to the previous thread,
the state of the system is defined as a reduced density $\rho\Ss = \tr\Sb \rho(t)$ where $\tr\Sb$ traces out the degrees of freedom of the environments. The theory of OQS seeks a completely positive operation $\mathcal{M}(t;t_i)$ 
(dynamical map, or quantum map) defined by\cite{Alicki2010}
\begin{equation}\label{eq:map-rho}
\rho\Ss(t) = \tr\Sb \{ U(t;t_i) \rho(t_i) U^\dagger(t;t_i) \} \equiv \mathcal{M}(t,t_i) \rho\Ss(t_i) 
\end{equation}
Alternatively, a quantum map can be considered as the given,
and then one of the many possible couplings to an environment
giving rise to same map after tracing out the environment
is called an environmental representation of that 
map~\cite{BengtssonZyczkowski}.
In either case, once the map is obtained, we can evaluate any quantity associated with the system.  For example
the transition probability from an initial pure state of the system $\ket{i}$ to a final pure state of the system $\ket{f}$ is given by
\begin{equation}\label{eq:P_if}
P_{if} = \tr\Ss \{\dyad{f} \mathcal{M}(t_i,t_f) \left(\dyad{i}\right) \} .
\end{equation}

In the path integral approach  
the dynamical map (\ref{eq:map-rho})
is written in coordinate basis as
\begin{equation}\label{eq:map-rho-Q}
\rho\Ss(Q_f,\tilde{Q}_f;t) = \iint \dd{Q_i} \dd{\tilde{Q}_i}\mathcal{M}(Q_f,\tilde{Q}_f;Q_i,\tilde{Q}_i;t_f,t_i) \rho\Ss(Q_i,\tilde{Q}_i,t_i)
\end{equation}
where $Q$ and $\tilde{Q}$ are coordinates of the system at the initial time $t_i$ and final time $t_{f}$. Feynman and Vernon wrote $\mathcal{M}$ in a path integral form
\begin{equation}\label{eq:map-rho-path}
\mathcal{M}(Q_f,\tilde{Q}_f;Q_i,\tilde{Q}_i;t_f,t_i) = \iint \mathcal{D}Q\, \mathcal{D}\tilde{Q}\, e^{i S[Q]}\, F(Q,\tilde{Q};t_f,t_i)\, e^{-i S[\tilde{Q}]}
\end{equation}
where $S[Q]$ is the classical action of a system trajectory without interactions
with the environment, and $Q(t)$ and $\tilde{Q}(t)$ are forward and time-reversed trajectories.  The effects of the environment are fully included in the \emph{influence functional} $F[Q,Q';t_f,t_i]$~\cite{Feynman1963,Feynman1965}.  
General properties of influence functionals were discussed 
in~\cite{Feynman1963}, and we will return to those below.

Exact expressions can be obtained when the 
functional integrals over the environment variables can be done in closed form.
In practice this means that integrals have to be Gaussian,
which is the case when the environment
is a Bose gas with Hamiltonian $H\Sb = \sum_k \left ( p_k^2/2 m_k+ m_k \omega_k^2 q_k^2/2 \right)$ initially in Gibbs states $\rho\Sb(t_i) = e^{-\beta H\Sb}/Z\Sb$, and
the coupling takes a bi-linear form $V\Ssb = X\Ss \otimes (\sum_x c_k q_k)$ where $X\Ss$ is an operator of the system, and $c_k$ is the coupling strength.
Such an environment is called a harmonic bath at inverse temperature $\beta$. 
We will later discuss the case when the environment consists
of two or more harmonic baths, each with its own temperature.

Under the above assumptions the influence functional can be written
in exponential form $F[Q,\tilde{Q};t_f,t_i] = e^{i\Phi[Q,\tilde{Q};t_f,t_i])}$ where
\begin{equation}\label{eq:influ-func-path}
\begin{split}
i \Phi[Q,\tilde{Q};t_f,t_i] &= i\int_{t_i}^{t_f} \dd{s} \int_{t_i}^{s} \dd{s'} \left [  (Q(s)-\tilde{Q}(s))(Q(s')+\tilde{Q}(s'))  \kappa_i\left(s,s'\right) \right ]\\
& \quad -\int_{t_i}^{t_f} \dd{s} \int_{t_i}^{s} \dd{s'} \left [ (Q(s)-\tilde{Q}(s))(Q(s')-\tilde{Q}(s'))\kappa_r\left(s,s'\right)\right ]
\end{split} 
\end{equation}
where $\Phi$ is known as the influence action or Feynman-Vernon action,
and $\kappa_r(\tau)$ and $\kappa_i (\tau)$ are known as dissipation and noise kernel.
These functions are
\begin{eqnarray}
\label{eq:kappa}
\kappa_i(s,s') &=& \sum_k \frac{c_k(s)c_k(s')}{2m_k\omega_k} \sin\omega_k(s-s') \\
\label{eq:k_r}
\kappa_r(s,s') &=& \sum_k \frac{c_k(s)c_k(s')}{2m_k\omega_k} \coth\left(\frac{\omega_k\beta}{2}\right)\cos\omega_k(s-s')
\end{eqnarray}
Except close to the initial and final times the time-dependence of
the coupling coefficients $c_k$ can be ignored, and the kernels
$\kappa_i$ and $\kappa_r$ then only depend on the time difference $\tau=s-s'$.

In the super-operator approach one instead directly evaluates Eq. (\ref{eq:full-evolution}).  Rewriting Eq. (\ref{eq:full-evolution}) using the Liouville operator 
$\mathcal{L}(t) \bullet= -i \comm{V\Ssb(t)}{\bullet}$ in the interaction picture, the map (in the interaction picture) can be written as
\begin{equation}\label{eq:map-rho-op}
\mathcal{M}(t_f,t_i) = \tr\Sb\left\{ \overleftarrow{\mathcal{T}} \exp \left[ \int_{t_i}^{t_f}\mathcal{L}(s)\dd{s}\right] (I \otimes \rho\Sb(t_i) )\right\}
\end{equation}
where $\overleftarrow{\mathcal{T}}$ is time-ordering super-operator which chronologically orders the super-operators (see Appendix \ref{sec:time-ordering-operator}.)
The symbol $\bullet$ in above and in the following is the ``slot'' on which
the super-operator acts, and represents any operator, including density operator. 
Using Wick's theorem, we find the map $\mathcal{M}(t_f,t_i) =  \overleftarrow{\mathcal{T}} e^{i\Phi(t_f,t_i)}$ with super-operator
\begin{equation}
\begin{split}
i \Phi(t_f,t_i) &=  i \int_{t_i}^{t_f} \dd{s}  \int_{t_i}^{s} \dd{s'} \left [ \left(\mathcal{X}\Ss^{+}(s)+\mathcal{X}\Ss^{-}(s)\right)\left(\mathcal{X}\Ss^{+}(s')-\mathcal{X}\Ss^{-}(s')\right)\kappa_i(s-s') \right] \\
& -\int_{t_i}^{t_f} \dd{s}  \int_{t_i}^{s} \dd{s'}\left[\left(\mathcal{X}\Ss^{+}(s)+\mathcal{X}\Ss^{-}(s)\right)\left(\mathcal{X}\Ss^{+}(s')+\mathcal{X}\Ss^{-}(s')\right)\kappa_r(s-s') \right ]
\end{split}  
\label{eq:inful-func-op}
\end{equation}
where $\mathcal{X}^{+}\Ss(t) \bullet = X\Ss(t) \bullet$ and $\mathcal{X}^{-}\Ss(t) \bullet =  -\bullet X\Ss(t)$ 
The two super-operators together can be expressed with commutator and anti-commutator as $(\mathcal{X}^{+}\Ss\pm \mathcal{X}^{-}\Ss) \bullet = [X\Ss,\bullet]_{\mp}$.
The kernels $\kappa_r$ and $\kappa_i$ in
\eqref{eq:inful-func-op}
are the same as those in Eq. \eqref{eq:influ-func-path}, and will be shown to be the
equilibrium pair correlation functions of the ideal Bose gas.

While the two methods use different mathematical objects, one with paths $Q(t)$ and $\tilde{Q}(t)$, and the other with super-operators $\mathcal{X}^{+}\Ss(t)$ and $\mathcal{X}^{-}\Ss(t)$, Eqs. (\ref{eq:influ-func-path}) and (\ref{eq:inful-func-op}) clearly show similarity.  They are the same if two quantities are replaced as $Q(t) \leftrightarrow  \mathcal{X}^{+}\Ss(t)$ and $\tilde{Q}(t) \leftrightarrow - \mathcal{X}^{-}\Ss(t)$.


Extension of these methods to a system interacting with multiple environments is straight-forward if the environments do not interact between
themselves. Indeed, \textit{General property of influence functionals 2}
of Feynman and Vernon states that
``If a number of [environments] act 
on [the system] and if $F^{k}$ is the influence of the $k$'th [environment] alone, 
then the total influence of all [the environments] 
is given by the product of the individual influences''~\cite{Feynman1963}.
In the super-operator approach the same statement
follows from the observation that if the Liouville operator
is a sum, say $\mathcal{L}(t) \bullet= -i \comm{V\Ssh(t)}{\bullet}
-i\comm{V\Ssc(t)}{\bullet}$,
and if the environment operators in
$V\Ssh(t)$ and $V\Ssc(t)$ commute and act on parts of the environment
that start in a product state (different baths),
then the time ordering of environment operators in \eqref{eq:map-rho-op}
can be done separately.


\section{Generating Function of Heat}
\label{sec:generating-function}
Once the dynamical map is found, we know the state of the system precisely.  However, the information on the state of environments is completely buried in the map.  If we want to investigate any quantity associated with the environments or correlation between the system and environments, the knowledge of the system density alone is not enough.
In order to make our story concrete, we consider a system interacting with a hot and a cold bath.  Their Hamiltonians are denoted as $H\Ss$, $H\Sh$, and $H\Sc$, respectively, and the interaction Hamiltonians between the system and the baths are  $V\Ssh$ and $V\Ssc$. 
As is well known, for harmonic baths the interaction Hamiltonian are accompanied by the Caldeira-Leggett counter-terms~\cite{CaldeiraLeggett83a}
which redefine the system Hamiltonian $H\Ss$. 

The initial state of the whole system is assumed to be a product state $\rho(t_i) =\rho\Ss(t_i) \otimes \rho\Sh(t_i) \otimes \rho\Sc(t_i)$ and the baths are at thermal equilibrium
\begin{equation}
   \rho_\ell(t_i) = \frac{1}{Z_\ell} \sum_n \ket{E_\ell(n)}e^{-\beta_\ell E_\ell(n)} \bra{E_\ell(n)},\qquad \ell=\textsc{h}, \textsc{c}
\end{equation}
where $Z_\ell$ is a partition function.  $E_\ell(n)$ and $\ket{E_\ell(n)}$ are eigenvalue and the corresponding eigenket of $H_\ell$. The system is initially in an arbitrary state $\rho_s(t_i) = \sum_i \mu_i \dyad{i}$ where $\mu_i$ and $\ket{i}$ are eigenvalues and eigenkets of the density.  

Now we want know the change in the energy of the cold bath, $\Delta E\Sc$, over time period $t_f-t_i$.
The probability distribution of $\Delta E\Sc$ may be written as 
\begin{equation}
P(\Delta E\Sc) = \sum_f\sum_i \mu_i P_{if}(\Delta E\Sc)
\end{equation}
where
\begin{equation}
\begin{split}
   P_{if}(\Delta E\Sc;t_f) &= \sum_{m,m'} \sum_{n,n'} |\bra{f,m',n'} U(t_f,t_i) \ket{i,m,n}|^2 \delta(\Delta E\Sc-E\Sc(n')+E\Sc(n))\\
   & \qquad \times \frac{e^{-\beta\Sh E\Sh(m)}}{Z\Sh} \frac{e^{-\beta\Sc E\Sc(n)}}{Z\Sc}
\end{split}
\end{equation}
$\mu_i$ and $\ket{i}$ are the eigenvalue and eigenket of $\rho\Ss(t_i)$ and the final states $\ket{f}$ can be any basis set.

The generating function of heat is the Fourier transform 
of parameter $\nu$ of the 
probability distribution $P(\Delta E\Sc;t_f)$ with respect to variable $\Delta E\Sc$.
One finds
\begin{eqnarray}\label{eq:G}
   G(\nu;t_f) &=& \int_{-\infty}^{\infty} P(q,t_f) e^{ i \nu q} \dd{q} \nonumber \\
   &=& \tr\Ss \Gamma(\nu;t_f)
\end{eqnarray}
where
\begin{equation}\label{eq:Gamma}
  \Gamma(\nu,t_f) = \tr\Sh\tr\Sc \left\{ e^{i \nu H\Sc} U(t_f,t_i) e^{-i \nu H\Sc}\rho(0) U^\dagger(t_f,t_i) \right\}.
\end{equation}
is an operator in the Hilbert space of the system. Direct comparison of Eqs (\ref{eq:map-rho}) and (\ref{eq:Gamma}) shows that $\Gamma(\nu,t)$ is quite similar to  $\rho\Ss(t)$. In fact, when $\nu=0$, they coincide.
The only difference is that one of the time evolution operators in Eq. (\ref{eq:Gamma}) is rotated by $e^{i \nu H\Sc}$.  

The resemblance suggests that the generating function can be computed with the methods developed for OQS.  Following the procedure discussed in the previous section, we first write $\Gamma$ with a map as
\begin{equation}\label{eq:map-Gamma}
\Gamma(\nu,t_f) = \mathcal{M}(\nu;t_f,t_i) \rho\Ss(t_i).
\end{equation}
where
%
%
%
\begin{equation}\label{eq:map-G-general}
\mathcal{M}(\nu;t_f,t_i) = \tr\Sh \tr\Sc \left \{ \left(I\Ss \otimes I\Sh \otimes e^{i \nu H\Sc}\right ) \exp\left [  \int_{t_i}^{t_f} \mathcal{L}(s) \dd{s} \right ]  \left( I\Ss \otimes \rho\Sh(t_i)  \otimes e^{-i \nu H\Sc} \rho\Sc(t_i) \right) \right\}
\end{equation}
We generally assume that the system interacts separately with the
hot and the cold bath, and thus the Liouville operator is split to two parts, $\mathcal{L}_\textsc{x} = -i \comm{V_\textsc{sx}(t)}{\cdot},\,\, \textsc{x}=\textsc{c}, \textsc{h}$. If the system parts of the two operators
(in the interaction picture) always commute
the expression further factorizes into the product of the traces
over the two baths separately.

The energy change associated with a particular transition from $\ket{i}$ to $\ket{f}$ can be expressed like the transition probability (\ref{eq:P_if}):
\begin{equation}\label{eq:G_if}
G_{if}(\nu,t) = \tr\Ss\{\dyad{f}\mathcal{M}(\nu;t_f,t_0)\left(\dyad{i}\right) \}
\end{equation}
    
In order to use the path integral approach, we express Eq. (\ref{eq:map-Gamma}) in the coordinate representation in the same way as Eq. (\ref{eq:map-rho-Q}),
\begin{equation}
\Gamma(\nu;Q_f,\tilde{Q}_f,;t_f,t_i)=\iint \dd{Q_i} \dd{\tilde{Q}_i}\mathcal{M}(\nu;Q_f,\tilde{Q}_f;Q_i,\tilde{Q}_i;t_f,t_i) \rho\Ss(Q_i,\tilde{Q}_i,t_i).
\end{equation}
The map has been derived using the path integral\cite{Aurell2018b} for the baths of ideal Bose gases and expressed with a new influence functional $F[\nu;Q,\tilde{Q};t_f,t_i]=e^{i\Phi\Sh[Q,\tilde{Q};t_f,t_i] +i \Phi\Sc[\nu;Q,\tilde{Q};t_f,t_i]}$. For the hot bath, $\Phi\Sh$ remains exactly the same as Eq. (\ref{eq:influ-func-path}) but for the cold bath, $\Phi\Sc$ is slightly changed to
\begin{equation}\label{eq:influ-func-G-path}
\begin{split}
i \Phi\Sc[\nu;Q,\tilde{Q};t_f,t_i] = &
-\int_{t_i}^{t_f} \dd{s} \int_{t_i}^{s} \dd{s'} 
\left[ Q(s)  Q(s') + \tilde{Q}(s)\tilde{Q}(s') \right ]\kappa_r(s-s') \\
&+ i \int_{t_i}^{t_f} \dd{s} \int_{t_i}^{s} \dd{s'} \left [ Q(s)  Q(s') - \tilde{Q}(s)\tilde{Q}(s') \right ]  \kappa_i(s-s') \\
&+ \int_{t_i}^{t_f} \dd{s} \int_{t_i}^{t_f} Q(s)\tilde{Q}(s') \left [ \kappa_r(s-s'+\nu) + i \kappa_i(s-s'+\nu) \right ] .
\end{split}
\end{equation} 
where for simplicity we write the kernels as they are away from the initial and final times.
When $\nu=0$, Eq. \eqref{eq:influ-func-G-path} is back to Eq. \eqref{eq:influ-func-path}.  The time shift $\nu$ in the cross correlation between forward and backward trajectories  contains all information about $\Delta E\Sc$.
In the path integral approach
\eqref{eq:influ-func-G-path}
emerges from rather complicated intermediate results
after cancellations and using
properties of hyperbolic and trigonometric functions.
For completeness we provide in
Appendix~\ref{app:path-integrals-harmonic-baths} an outline of these
results previously announced in ~\cite{Aurell2018-Erratum}.

Based on the correspondence between the path integral method and the super operator method, we expect that the map defined in Eq. (\ref{eq:map-G-general}) is given by $\mathcal{M}(\nu;t_f,t_i) = e^{i\Phi\Sh(t_f,t_i) +i\Phi\Sc(\nu;t_f,t_i)}$ with
\begin{equation}\label{eq:influ-func-G-op}
\begin{split}
i \Phi\Sc(\nu;t_f,t_i) = & -\int_{t_i}^{t_f} \dd{s} \int_{t_i}^{s} \dd{s'} 
\left[ \mathcal{X}^{+}(s)  \mathcal{X}^{+}(s') + \mathcal{X}^{-}(s)\mathcal{X}^{-}(s') \right ]\kappa_r(s-s') \\
& + i \int_{t_i}^{t_f} \dd{s} \int_{t_i}^{s} \dd{s'} \left [\mathcal{X}^{+}(s)  \mathcal{X}^{+}(s') - \mathcal{X}^{-}(s)\mathcal{X}^{-}(s') \right ]  \kappa_i(s-s')\\
&- \int_{t_i}^{t_f} \dd{s} \int_{t_i}^{t_f} \mathcal{X}^{+}(s)\mathcal{X}^{-}(s')x \left [\kappa_r(s-s'+\nu)+i \kappa_i(s-s'+\nu) \right]
\end{split}
\end{equation}
$\Phi\Sh(t_f;t_i)$ remains the same as Eq. (\ref{eq:inful-func-op}).  In the next section, we derive Eq. \eqref{eq:influ-func-G-op},
and show that it appears more directly in the super-operator approach.

\section{Super-operator Approach for the Generating Function}
\label{sec:general-theory}
We derive Eq. \eqref{eq:influ-func-G-op}  by evaluating the super-operator expression of map
\begin{equation}\label{eq:map_c-G}
\mathcal{M}\Sc(\nu;t_f,t_i) = \tr\Sc \left\{e^{i \nu H\Sc}\exp\left[ \int_{t_i}^{t_f} \mathcal{L}\Sc(s) \dd{s}\right]I\Ss\otimes e^{-i H\Sc} \rho\Sc(t_i) \right\} .
\end{equation}  
For simplicity we have assumed that the system parts
of the interaction Hamiltonians commute, which allow
us to focus on the trace over the cold bath.

First we rewrite $H\Sb$ with creation and annihilation operators, $a^\dagger_k$ and $a_k$: 
\begin{equation}\label{eq:H_B}
    H\Sb = \sum_k \omega_k a_k^\dagger a_k
\end{equation}
and the interaction Hamiltonian $V\Sc = X\Ss \otimes Y\Sc$ with
\begin{equation}\label{eq:YB}
    Y\Sc = \sum_k c'_k (a_k^\dagger + a_k)
\end{equation}
where $c'_k= c_k/\sqrt{2 m_k \omega_k}$ is coupling strength. The system part of the coupling $X_s$ is arbitrary.

Using the interaction picture, the Liouville super-operator is defined by
\begin{equation}
	\mathcal{L}\Sc(t)\, \bullet = -i \comm{X\Ss(t)\otimes Y\Sb(t)}{\bullet} = -i \sum_{d=\pm} \left\{\mathcal{X}\Ss^d (t)\otimes \mathcal{Y}\Sc^d(t)\right \} \, \bullet
\end{equation}
where $X\Ss(t) = e^{i H\Ss (t-t_i)} X\Ss e^{-i H\Ss (t-t_i)}$ and $e^{iH\Sb (t-t_i)} Y\Sb ^{-i H\Sb (t-t_i)}$ and for mathematical convenience, we introduced the following super-operators
\begin{equation}
\label{eq:super-operator-plus-minus-def}
\mathcal{X}\Ss^{+}\, \bullet = X\Ss\, \bullet, \quad \mathcal{X}\Ss^{-} \, \bullet = - \bullet\, X\Ss, \quad  
\mathcal{Y}\Sc^{+}\, \bullet = Y\Sc\, \bullet, \quad \mathcal{Y}\Sc^{-}\, \bullet =  \bullet\, Y\Sc.
\end{equation}

Expanding the exponential function in Eq. (\ref{eq:map_c-G})

\begin{multline}\label{eq:dynamical_map}
     \mathcal{M}_c(\nu;t_f,t_i) 
     = \sum_n  \frac{1}{n!} \int_{t_i}^{t_f} \dd{t_1} \cdots \int_{t_i}^{t_f}\dd{t_n} \tr\Sc \left \{e^{i \nu H\Sc}  \overleftarrow{\mathcal{T}} \mathcal{L}(t_1) \cdots \mathcal{L}(t_n)  \left ( I\Ss \otimes e^{-i \nu H\Sc} \rho\Sb(t_i) \right ) \right \} \\
     =   \sum_n \frac{(-i)^n}{n!} \int_{t_i}^{t_f} \dd{t_1} \cdots \int_{t_i}^{t_f} \dd{t_n}  \sum_{d_1} \cdots \sum_{d_n} \overleftarrow{\mathcal{T}} \left ( \mathcal{X}\Ss^{d_1}(t_1) \cdots \mathcal{X}\Ss^{d_n}(t_n)\right ) \\
       \times C^{d_1,\cdots,d_n}(\nu;t_1,\cdots,t_n)
\end{multline}
where multi-time correlation functions of the environment are defined as
\begin{align}\label{eq:multi-time-correlation}
C^{\nu;d_1,\cdots,d_n}(t_1,\cdots,t_n) &=   \tr\Sc \left [ \overleftarrow{\mathcal{T}}\left \{ e^{i\nu H\Sc}\mathcal{Y}\Sc^{d_1} (t_1) \cdots \mathcal{Y}\Sc^{d_n} (t_n) e^{-i \nu H\Sc}\right \} \rho\Sc(t_1) \right ] \nonumber\\
&= \expval{ \overleftarrow{\mathcal{T}}  [e^{i \nu H\Sc} \mathcal{Y}\Sc^{d_1} (t_1)e^{-i \nu H\Sc}] \cdots [e^{i \nu H\Sc} \mathcal{Y}\Sc^{d_n} (t_n)e^{-i \nu H\Sc}]  }_{t_i}
\end{align}
where $\expval{\cdots}_{t_i}$ indicates expectation value $\tr\Sc\{ \cdots \rho\Sc(t_i)\}$.
Since $H\Sc$ is quadratic in $a$ and $a^\dagger$, all odd order correlation functions vanish.  For the even order terms, we apply the Wick's theorem for operators $e^{i \nu H\Sc} \mathcal{Y}\Sc^{d_1} (t)e^{-i \nu H\Sc}$
\begin{equation}
    C^{d_1, \cdots d_{2n}}(t_1, \cdots, t_{2n}) 
    =\sum_{\substack{\text{all possible}\\ \text{pairing}}} \quad \prod_{\text{all pairs}} C^{d_j,d_k} (t_j,t_k)
\label{eq:Wick-theorem}
\end{equation}
where $\sum_{\text{all pairing}}$ indicates the sum of all possible combinations of pairs.  The map is now expressed with the pair correlation functions as
\begin{align}
\mathcal{M}\Sc(\nu;t_f,t_i) &= \overleftarrow{\mathcal{T}}
 \sum_n \frac{(-i)^{2n}}{2^n n!} \left[\int_{t_i}^{t_f} \dd{t_1} \int_{t_i}^{t_f} \dd{t_2}  \sum_{d_1}  \sum_{d_2} \mathcal{X}\Ss^{d_1}(t_1) \mathcal{X}\Ss^{d_2}(t_2) C^{d_1,d_2}(\nu;t_1,t_2) \right ]^n \nonumber \\
 &= \overleftarrow{\mathcal{T}} \exp \left [-\frac{1}{2} \int_{t_i}^{t_f} \dd{t_1} \int_{t_i}^{t_f} \dd{t_2} \sum_{d_1}  \sum_{d_2} \mathcal{X}\Ss^{d_1}(t_1) \mathcal{X}\Ss^{d_2}(t_2) C^{d_1,d_2}(\nu;t_1,t_2) \right ]\label{eq:map_after_Wick}
\end{align}

\begin{figure}
	\includegraphics[width=6in]{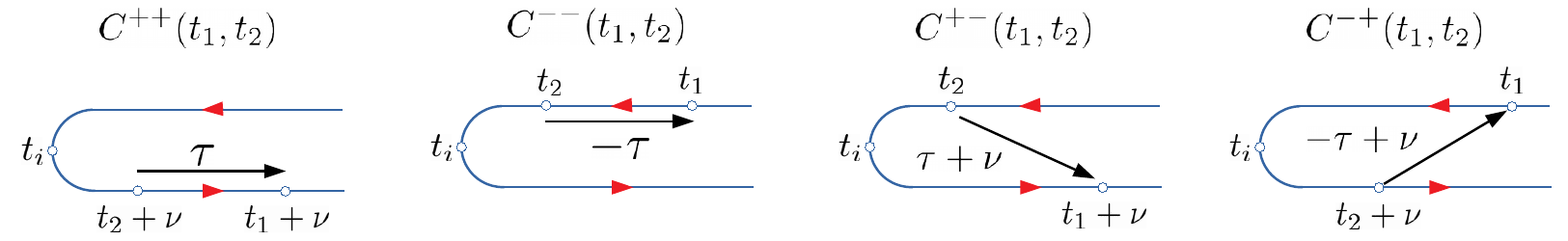}
	\caption{Time-line diagrams for two-time correlation for $\tau = t_1-t_2> 0$.  The upper (lower) branch shows anti-chronological (chronological) time line.  The time on chronological branch shifts by $\nu$.  For $C^{++}(t_1,t_2)$, both times shift by the same amount and thus the time difference is not affected by the shift.  The situation for $C^{--}(t_1,t_2)$ is identical to the normal correlation. For the cross correlation $C^{+-}(t_1,t_2)$, only $t_1$ on the chronological branch shifts and thus the time difference also shifts.  The situation for $C^{-+}(t_1,t_2)$  is similar to $C^{+-}(t_1,t_2)$ except that $t_2$ shifts by $\nu$ instead of $t_1$.  For $t_1<t_2$, the direction of $\tau$ is reversed for $C^{++}(t_1,t_2)$ and $C^{--}(t_1,t_2)$. However, the sign of $t_1-t_2$ does not affect the situation of $C^{+-}(t_1,t_2)$ and $C^{-+}(t_1,t_2)$.}\label{fig:2points-time-line}
\end{figure}
The four pair correlation functions $C^{++}$, $C^{+-}$, $C^{-+}$, and $C^{--}$  can be expressed with the standard pair correlation function $\kappa(\tau)$ as shown in Fig. \ref{fig:2points-time-line}. (See Appendix \ref{sec:pair-correlation}.)
The correlation functions between the two times on the same branch are 
	\begin{equation}
		C^{++}(t_1,t_2) = \kappa_r(\tau) - i \sign(\tau) \kappa_i(\tau),
		\qquad
		C^{--}(t_1,t_2) = \kappa_r(\tau) + i \sign(\tau) \kappa_i(\tau)
	\end{equation}
where $\tau=t_1-t_2$.  The cross correlation functions are
	\begin{equation}
		C^{+-}(t_1,t_2) = \kappa_r(\tau+\nu)+ i \kappa_i(\tau+\nu)
		\qquad
		C^{-+}(t_1,t_2) = \kappa_r(\tau-\nu)- i \kappa_i(\tau-\nu)
	\end{equation}
where we have again stated the form these kernels take away from the initial and final time.
Substituting these correlation functions into \eqref{eq:map_after_Wick} we obtain Eq. (\ref{eq:influ-func-G-op}).  Note that only the difference between the map for $\rho$ derived and discussed in Section~\ref{sec:OQS}
and the map for the generating function is the cross correlations.

\section{An-harmonic baths and cluster expansions}
\label{sec:non-harmonic}
A second advantage of the super-operator formulation 
is in the derivation of corrections to the Feynman-Vernon theory.
The starting point is then the dynamical map \eqref{eq:dynamical_map}
with the multi-time correlation functions of the environment \eqref{eq:multi-time-correlation},
but without assuming Wick's theorem. 
The outcome will be that multi-time cumulants of the environment (discussed below) 
translate into kernels of higher-than-quadratic contributions to the Feynman-Vernon action.

For ordinary operator correlation functions, successive orders of cumulants are defined inductively as
\begin{align}
G_1(t_1) &=    C(t_1) \nonumber \\
G_2(t_1,t_2) &=    C(t_1,t_2) -  G_1(t_1) G_1(t_2)  \nonumber \\ 
G_3(t_1,t_2,t_3) &=    C(t_1,t_2,t_3) -   G_1(t_1) G_1(t_2) G_1(t_3) - G_1(t_1) G_2(t_2,t_3)    \nonumber \\
&\qquad - G_1(t_2) G_2(t_1,t_3) -  G_1(t_3) G_2(t_1,t_2)     \nonumber \\ 
\label{eq:cumulants-123}
&\vdots   
\end{align}
Owing to the time-ordering super-operator $\overleftarrow{\mathcal{T}}$ and indexes $d_j$, $C^{d_1,\cdots,d_n}(t_1,\cdots,t_n)$ defined in Eq. \eqref{eq:multi-time-correlation} behaves like an ordinary multi-time correlation function and the relations \eqref{eq:cumulants-123} hold. Hence,
\begin{equation}
    C^{d_1, \cdots d_{N}}(t_1, \cdots, t_{N}) = \sum_{\substack{\text{all possible}\\ \text{groupings}}}\quad \prod_{\substack{\text{groups of}\\ \text{one time}}} G_1^{d_1} (t_1)
\prod_{\substack{\text{groups of} \\ \text{two times}}} G_2^{d_1,d_2} (t_2,t_3) \quad \cdots
\label{eq:cumulant-expansion}
\end{equation}
where $N$ can be even or odd.
The first order cumulant ($G^{d}_1$) can be set to zero by a shift. The first non-trivial cumulant is then
\begin{equation}\label{eq:cumulants-4}
\begin{split}
   G_4^{d_1,d_2,d_3,d_4}(t_1,t_2,t_3,t_4) &=    C^{d_1,d_2,d_3,d_4}(t_1,t_2,t_3,d_4) - C^{d_1,d_2}(t_1,t_2) C^{d_3,d_4}(t_3,t_4)  \\
& - C^{d_1,d_3}(t_1,t_3) C^{d_2,d_4}(t_2,t_4) 
- C^{d_1,d_4}(t_1,t_4) C^{d_2,d_3}(t_2,t_3) 
\end{split}
\end{equation}
where we have retained the super-operator notation on the right-hand side.
For a bath that satisfies Wick's theorem, this cumulant and all others beyond $G_2^{d_1,d_2}$ vanish.

The second step is to count the number of groupings
in \eqref{eq:cumulant-expansion} with $n_1$ groups of one element,
$n_2$ groups of two elements (pairs), $n_3$ groups of three elements, etc. 
There are $\displaystyle\frac{N!}{n_1!n_2!(2!)^{n_2}n_3!(3!)^{n_3}\cdots }$ such groupings.
The correlation functions appear inside the time integral and index sums 
in \eqref{eq:dynamical_map}
and the indices and time variables can therefore be renamed in any way.  
Each grouping of the same type (same $n_1,n_2,\ldots$)
hence contributes the same, and the quantum map 
can be summed in an analogous way to Section~\ref{sec:general-theory}.

Introducing $Q(t)$ as the coordinate representation of $\mathcal{X}\Ss^{+}(t)$
and  $-\tilde{Q}(t)$ the coordinate representation of $\mathcal{X}\Ss^{-}(t)$ 
one can show that the contribution to the 
Feynman-Vernon action from $n$ number of $X$ and $m$ number $Y$ is
\begin{equation}\label{eq:cumulants-5}
\begin{split}
S^{(n,m)}&= (-i)^{n} (i)^{m}\int_{t_0}^{t}\dd{s_1} \int_{t_0}^{s_1}  \dd{s_2}\cdots \int_{t_0}^{t}\dd{u_1} \int_{t_0}^{u_1}  \dd{u_2} \cdots
Q(s_1)Q(s_2)\cdots Q(s_n) \\
&\quad \times \tilde{Q}(u_1) \tilde{Q}(u_2)\cdots \tilde{Q}(u_m)  G_{n+m}(u_m,\ldots,u_1,s_1,\ldots,s_n)
\end{split}
\end{equation}
where the last term is the cumulant of the operator correlation function with the times
ordered as required in the super-operator cumulant. One can further sum all contributions of the same order
and express them in terms of time-ordered sums $\zeta_{+}(t)=Q(t)+\tilde{Q}(t)$ and differences $\zeta_{-}(t)=Q(t)-\tilde{Q}(t)$.
The most important general result one can find this way is 
for the largest time, the dependence in only through the difference $\zeta_{-}(t)$
as also follows from 
Feynman and Vernon's \textit{General property of influence functionals 5}~\cite{Feynman1963}.
Ultimately this is a consequence of the super-operator correlation function
$C^{d_1,\cdots,d_n}(t_1,\cdots,t_n)$ being independent of the symbol connected to the largest time.
Otherwise the general expressions are somewhat unwieldy, and we will here only quote
the result to third order
\begin{equation}
\label{eq:third-cumulant-main-text}
\begin{split}
& \sum_{n+m=3}S^{(n,m)} = \frac{i}{4} \int_{t_0}^{t} \dd{s} \zeta_{-}(s)\\
& \quad\times \int_{t_0}^{s} \dd{u} \int_{t_0}^{u} \dd{v}  \Big(\zeta_{+}(u) \zeta_{+}(v) A +\zeta_{+}(u)\zeta_{-}(v) B + \zeta_{-}(u) \zeta_{+}(v) C + \zeta_{-}(u) \zeta_{-}(v) D \Big)
\end{split}
\end{equation}
where $A$, $B$, $C$ and $D$ are combinations of third order bath correlation functions 
given in Appendix~\ref{app:time-ordered-cumulants}.

\section{Discussion}
\label{sec:discussion}
In this paper we have compared the path integral and super-operator
approaches to the theory of open quantum system (OQS).
We have pointed out that both approaches lead to 
equivalent descriptions of a system interacting with one
or several harmonic oscillator baths, but that the routes to the result 
are qualitatively different.
In the super-operator approach the kernels in the description 
are found to be certain pair correlation functions of the bath (or baths),
and the main assumption is Wick's theorem, reducing any 
correlation function to sums of products of pair correlation functions.
In the path integral approach, the 
result on the hand follow from integrating over the initial and final
points of the propagator of an harmonic oscillator (one of the degrees
of freedom of the bath) acted upon by a linear drive 
(a linear interaction with the system), and after a fair amount of cancellation.

We have here shown that same holds for the generating function of heat: both approaches give the same result, but the super-operator approach is more direct.
In particular, the fact that the generating function of heat 
can be expressed with the same kernels as for the system density matrix (Feynman-Vernon theory),
with only a time shift in the terms mixing the forward and time-revered paths, 
follows in a much more straight-forward manner in the super-operator approach.

We have also shown that the super-operator approach extends in a natural
way to interactions with environments 
where Wick's theorem does not hold.
Cumulants of correlation functions
of the environment, which vanish when Wick's theorem holds,
hence translate to kernels in higher-order terms in the Feynman-Vernon 
action.
In the text we have discussed that the resulting
higher-order theory of the influence functional
satisfies the general properties stated by Feynman and Vernon. 
Considerations of when the higher-order terms are
comparable or 
more important than the Feynman-Vernon terms 
are left for future work.

Several of the results in this paper can be found in the literature
and it is therefore appropriate to discuss antecedents.
The super-operator expression for evolution operator of the reduced
density matrix of the system (Eq.~\eqref{eq:inful-func-op} above)
is given (in the Schr\"odinger picture) as Eq. (3.508) on page 187 in the monograph of Breuer and
Petruccione~\cite{Breuer2002}.
Two recent contributions that use a similar plus/minus (left/right)
representation of the super-operator as we do are 
\cite{DiosiFerialdi2014} and
\cite{GasbarriFerialdi2018}; the latter paper also 
extends the analysis beyond harmonic baths, though in a different manner than we do.
Time shifts in kernels describing a statistics of heat appear
in the theory of heat transport through a Josephson junction
developed in~\cite{Golubev2013}, though in a particular setting,
and for a partially classical model. 
We have here strived to gather together these earlier results in a coherent
whole, and in the context of current concerns in quantum thermodynamics.

We end by summarize the assumptions that go 
and do not go into the new higher-order theory we have developed here.
First, we assume that 
the system and the environment start out in 
a product state. Second, we assume that it is possible 
to write the system-environment interaction as
$V\Ssb = \sum_k X\Ss^{k} \otimes B\Sb^{k}$, where $X\Ss^{k}$
and $B\Sb^{k}$ are operators on respectively the system and
the environment, and where all the $B\Sb^{k}$ commute.
Third, we assume that the initial state 
of the environment is 
a product state compatible with the interaction.
By the latter we mean that if 
the full environment Hilbert space is a product space
$\mathcal{H}\Sb = \mathcal{H}\Sb^{1}\otimes \mathcal{H}\Sb^{2}\cdots$
and the operators $B\Sb^{j}$ act on $\mathcal{H}\Sb^{j}$,
then 
the initial environment density matrix factorizes as as  
$\rho\Sb = \rho\Sb^{1}\otimes \rho\Sb^{2}\cdots$ where 
$\rho\Sb^{j}$ is a unit trace positive Hermitian operator on
$\mathcal{H}\Sb^{j}$.
One class of models that fulfill the above is when the system
interacts with one or several baths which start out independent, and
which do not interact between themselves.
In the other direction, in each bath the environmental degrees of freedom can 
be either Bosonic or Fermionic (or both), and the Hamiltonians can be arbitrary.
The initial state of each bath does not even have to be in equilibrium.
We suspect that such a general-looking result will find applications
also outside the current realm of theory of open quantum system. 

\ack

This work was initiated at the Nordita program
``New Directions in Quantum Information'' (Stockholm, April 2019).
We thank Nordita, Quantum Technology Finland (Espoo, Finland), 
and International Centre for Theory of Quantum Technologies
(Gda\'nsk, Poland)
for their financial support for this event.
EA thanks Dr Dmitry Golubev for discussions.  RK thanks Garrett Higginbotham and Saarth Anjali Chitale for helpful discussion.

\appendix

\section{Unitary time evolution of a density operator and time-ordering super-operator}\label{sec:time-ordering-operator}
We consider first unitary time-evolution of a ket $\ket{\psi(t)}$ and a bra $\bra{\psi(t)}$ under a Hamiltonian $H=H_0+V$ where $H_0$ is an unperturbed Hamiltonian and $V$ a perturbation. Using the interaction picture $V(t) = e^{i H_0 t} V e^{-i H_0 t}$ the time evolution of the ket and bra can be expressed with a time evolution operator.
\begin{equation}
\ket{\psi(t)} = U(t,t_i) \ket{\psi(t_i)}, \qquad \bra{\psi(t)} = \bra{\psi(t_i)} U^\dagger(t,t_i)
\end{equation}
where the forward and backward evolution operators are defined by
\begin{subequations}
\begin{equation}
U(t,t_i) = \overleftarrow{T} \exp\left [ - i \int_{t_i}^t V(s) \dd{s}\right]
\end{equation}
\begin{equation}
U^\dagger(t,t_i) = \overrightarrow{T} \exp\left [ i \int_{t_i}^t V(s) \dd{s}\right]
\end{equation}
\end{subequations}
where $\overleftarrow{T}$ and $\overrightarrow{T}$ are chronological and anti-chronological time ordering operator.

\begin{figure}
	\center
	\includegraphics[width=3in]{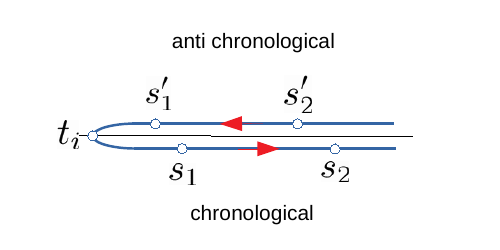}
	\caption{Time evolution of density operator, the operators $\widetilde{V}(s)$ are ordered along the time line from $t$ to $t_i$ (anti-chronological order) and then from $t_i$ to $t$ (chronological order). This particular diagram shows the case of $V(s_2) V(s_1) \rho(t_i) V(s'_1) V(s'_2)$. }
	\label{fig:time-line}
\end{figure}

The evolution of a density operator involves both forward and backward evolution operators as
\begin{equation}
\rho(t)=U(t,t_i) \rho(t_i) U^\dagger(t,t_i).
\end{equation} 
Managing the order of operators is a bit complicated due to the presence of two evolutions.  There is a simpler expression using the time line shown in Fig. \ref{fig:time-line}. We note that the evolution of the density operator is determined by the Liouville-von Neumann equation
\begin{equation}
\dv{\rho(t)}{t} = \mathcal{L}(t) \rho(t)
\end{equation}
where the Liouville super-operator is defined by $\mathcal{L}(t) = -i \comm{V(t)}{\bullet}$. Then,
\begin{equation}
\rho(t) = \overleftarrow{\mathcal{T}} \exp \left[\int_{t_i}^t \mathcal{L}(s)\dd{s}\right] \rho(t_i)
\end{equation}
where the time-ordering super-operator $\overleftarrow{\mathcal{T}}$ orders super-operators such as $\mathcal{L}(t)$ chronologically. It automatically orders regular operators along the time line shown in Fig. \ref{fig:time-line}.  As an example, consider $t_1>t_2$, 
\begin{align}
&\overleftarrow{\mathcal{T}} \mathcal{L}(t_2) \mathcal{L}(t_1) \rho(t_i)
=  \mathcal{L}(t_1) \mathcal{L}(t_2) \rho(t_i) \nonumber\\
& \qquad =   V(t_1)V(t_2)\rho(t_i)-V(t_1)\rho(t_i)V(t_2) - V(t_2)\rho(t_i)V(t_1) + \rho(t_i) V(t_2) V(t_1)
\end{align}
which automatically orders $V(t)$ chronologically if it is on the left of $\rho(t_i)$ and anti-chronologically on the right.

\section{Pair correlation functions}\label{sec:pair-correlation}

Now, we evaluate the four pair correlation functions $C^{++}$, $C^{+-}$, $C^{-+}$, and $C^{--}$ and express them with an ordinary  correlation function 
\begin{equation}
C(t_1,t_2) = \expval{Y\Sc(t_1) Y\Sc(t_2) \rho\Sc(t_i)} = \kappa_r(t_1-t_2) - i \kappa_i(t_1-t_2)
\end{equation}
where $\kappa_r(\tau)$ and $\kappa_i(\tau)$ are shown in Eq. \eqref{eq:kappa}.

For the diagonal ones, we find the exactly the same correlation functions as those in the influential function as follows:
\begin{align}
C^{++}(t_1,t_2) &= \tr\Sc \left \{ \overleftarrow{\mathcal{T}}  e^{i \nu H\Sc} \mathcal{Y}\Sc^{+}(t_1)  \mathcal{Y}\Sc^{+}(t_2) e^{-i \nu H\Sc} \rho\Sc(t_i) \right \}  \nonumber\\
&= 
\begin{cases} \tr\Sc \left \{  e^{i \nu H\Sc} Y\Sc(t_1) Y\Sc(t_2)e^{-i \nu H\Sc}  \rho\Sc(t_i) \right \} & t_1>t_2 \\
\tr\Sc \left \{  e^{i \nu H\Sc} Y\Sc(t_2) Y\Sc(t_1)e^{-i \nu H\Sc}  \rho\Sc(t_i) \right \} & t_1<t_2
\end{cases}  \nonumber\\
&=  
\begin{cases}
\expval{Y\Sc(t_1) Y\Sb(t_2)} = C(t_1,t_2)  & t_1>t_2 \\
\expval{Y\Sc(t_2) Y\Sb(t_1)} = C^*(t_1,t_2)  & t_1<t_2  
\end{cases}
\end{align}
\begin{align}
C^{--}(t_1,t_2) &= \tr\Sc \left \{ \overleftarrow{\mathcal{T}}  e^{i \nu H\Sc} Y\Sc^{<}(t_1)  Y\Sc^{<}(t_2) e^{-i \nu H\Sc}  \rho\Sc(t_i) \right \}  \nonumber\\
&= 
\begin{cases} \tr\Sc \left \{  \rho\Sc(t_i)  Y\Sc(t_2) Y\Sc(t_1) \right \} & t_1>t_2 \\
\tr\Sc \left \{  \rho\Sc(t_i)  Y\Sc(t_1) Y\Sc(t_2) \right \} & t_1<t_2 
\end{cases}  \nonumber\\
&=  
\begin{cases}
\expval{Y\Sc(t_2) Y\Sc(t_1) } = C(t_1,t_2)^*  & t_1>t_2 \\
\expval{Y\Sc(t_1) Y\Sc(t_2) } = C(t_1,t_2)  & t_1<t_2  
\end{cases} 
\end{align}
where we used $\comm{e^{i \nu H\Sc}}{\rho\Sc(t_i)}=0$.
A standard correlation function $C(t_1,t_2) = \expval{Y\Sb(t_1) Y\Sb(t_2)}$ and its complex conjugate $C^*(t_1,t_2) = \expval{Y\Sb(t_2) Y\Sb(t_1)}$ are used in the final expression.  Notice that these two correlation functions are exactly the same as ones in the influence functional.

However, the off-diagonal ones are different.
\begin{align}
C^{+-}(t_1,t_2) &=    \tr\Sc \left \{ \overleftarrow{\mathcal{T}}  e^{i \nu H\Sc} \mathcal{Y}\Sc^{>}(t_1)  \mathcal{Y}\Sc^{<}(t_2) e^{-i \nu H\Sc}  \rho\Sc(t_i) \right \}  \nonumber\\
&= \begin{cases}
\tr\Sc \left \{ e^{i \nu H\Sc} \mathcal{Y}\Sc^{>}(t_1)  \mathcal{Y}\Sc^{<}(t_2) e^{-i \nu H\Sc}  \rho\Sc(t_i) \right \}  & t_1>t_2\\
\tr\Sc \left \{ e^{i \nu H\Sc} \mathcal{Y}\Sc^{<}(t_2)  \mathcal{Y}\Sc^{>}(t_1) e^{-i \nu H\Sc}  \rho\Sc(t_i) \right \}  & t_2>t_1
\end{cases}\nonumber \\
&= 
\tr\Sc \left \{  e^{i \nu H\Sc} Y\Sc(t_1)e^{-i \nu H\Sc}\,   \rho\Sc(t_i)\,  Y\Sc(t_2) \right \}  \nonumber\\
&=  
\expval{ Y\Sc(t_2)Y\Sc(t_1+\nu) } =  C(t_2,t_1+\nu)
\end{align}
\begin{align}
C^{-+}(t_1,t_2) &= \tr\Sc \left \{ \overleftarrow{\mathcal{T}}  e^{i \nu H\Sc} \mathcal{Y}\Sc^{<}(t_1) \mathcal{Y}\Sc^{>}(t_2) e^{-i \nu H\Sc} \rho\Sc(t_i) \right \}  \nonumber\\
&= \begin{cases}
\tr\Sc \left \{ e^{i \nu H\Sc} \mathcal{Y}\Sc^{<}(t_1)  \mathcal{Y}\Sc^{>}(t_2) e^{-i \nu H\Sc}  \rho\Sc(t_i) \right \}  & t_1>t_2\\
\tr\Sc \left \{ e^{i \nu H\Sc} \mathcal{Y}\Sc^{>}(t_2)  \mathcal{Y}\Sc^{<}(t_1) e^{-i \nu H\Sc}  \rho\Sc(t_i) \right \}  & t_2>t_1
\end{cases}\nonumber \\
&=  
\tr\Sc \left \{  e^{i \nu H\Sc} Y\Sc(t_2)e^{-i \nu H\Sc}\  \rho\Sc(t_i)\,  Y\Sc(t_1) \right \}
\nonumber\\
&=  
\expval{Y\Sc(t_1) Y\Sc(t_2+\nu)  } = C(t_1,t_2+\nu)
\end{align}
where time on the chronological branch shifts by $\nu$.

\section{Time-ordered cumulant expansion}
\label{app:time-ordered-cumulants}
The starting point is an expansion analogous to \eqref{eq:map_after_Wick}
but using the cumulant expansion \eqref{eq:cumulant-expansion}
instead of Wick's theorem.
Both even and odd terms may appear.
We can consider interchanges within one group, say\\
$G_3^{d_1,d_2,d_3}(t_1,t_2,t_3)G_3^{d_4,d_5,d_6}(t_4,t_5,t_6)$
with $G_3^{d_1,d_2,d_4}(t_1,t_2,t_4)G_3^{d_3,d_5,d_6}(t_3,t_5,t_6)$, which will contribute the same.
This means that the quantum map can now be simplified to
\begin{align}
    \mathcal{M}(t_f;t_0) &=  \sum_N \frac{(-i)^{N}}{N!} \sum_{n_1+2n_2+3n_3+\cdots=N}
\frac{N!}{n_1!n_2!(2!)^{n_2}n_3!(3!)^{n_3}\cdots }
\int_{t_i}^{t_f} \dd{t_1} \cdots \int_{t_i}^{t_f} \dd{t_N}  \nonumber \\
    & \hspace{0.5in} \times \sum_{d_1} \cdots \sum_{d_N} 
    \overleftarrow{\mathcal{T}}\left \{\mathcal{X}\Ss^{d_1}(t_1) \cdots \mathcal{X}\Ss^{d_{N}}(t_N)\right\} 
     \prod_i G_1^{d_i} (t_i) \prod_{ij} G_2^{d_i,d_j} (t_i,t_j) \cdots
\nonumber \\
    &= 
\overleftarrow{\mathcal{T}} \exp \left[ \frac{(-i)^2}{2!} \int_{t_0}^{t_f} \dd{t_1}  \int_{t_0}^{t} \dd{t_2}\sum_{d_1,d_2} 
G_2^{d_1,d_2}(t_1,t_2) \mathcal{X}\Ss^{d_1}(t_1)\mathcal{X}\Ss^{d_2}(t_2) \right. \nonumber \\
& + \frac{(-i)^3}{3!} \int_{t_i}^{t_f} \dd{t_1}  \int_{t_i}^{t_f} \dd{t_2} \int_{t_i}^{t_f} \dd{t_3}\sum_{d_1,d_2,d_3}
G_3^{d_1,d_2,d_3}(t_1,t_2,t_3) \mathcal{X}\Ss^{d_1}(t_1)\mathcal{X}\Ss^{d_2}(t_2)\mathcal{X}\Ss^{d_3}(t_3) \nonumber \\
& \left. + \cdots \right] \label{eq:map_without_Wick}
\end{align}
In the last equation we have used that the operators have zero mean, $G_1=0$.
The second order term is the standard Feynman-Vernon expansion as evaluated above.

The third term can be evaluated as follows.  The cumulants with the same indexes  $G_3^{+++}$ and $G_3^{---}$ remain the same when 
the times are permuted. This can be done in $3!$ different ways.  If only two of the indexes $d_i$ and $d_j$ are the same, e.x., $G_3^{++-}$, give a factor two if permuted, and this can be done in three different ways.
Furthermore $G_3^{++-}(t_1,t_2,t_3)$ equals $C(t_3,t_1,t_2)$ if $t_1>t_2$
and $C(t_3,t_2,t_1)$ if $t_1<t_2$ where $C$ is the operator correlation function,
and similarly for the other cases.
Introducing for convenience $Q_{t}$ the coordinate representation of $\mathcal{X}\Ss^{+}(t)$
and  $-\tilde{Q}_{t}$ the coordinate representation of $\mathcal{X}\Ss^{-}(t)$,
and $S^{(n,m}$ for the terms with $n$ number of $Q$ and $m$ number of $\tilde{Q}$,
the sum of all terms to third order is thus
\begin{equation}
\label{eq:third-cumulant}
\begin{split}
\sum_{n+m=3}S^{(n,m)} =& i\left[ \int_{t_i}^{t_f} \dd{s}  \int_{t_i}^{s} \dd{u} \int_{t_i}^{u} \dd{v}  Q(s) Q(u) Q(v) C(s,u,v) \right.\\
& \hspace{1.5in} \left. - \tilde{Q}(s) \tilde{Q}(u) \tilde{Q}(v) C(v,u,s) \right] \\
& +i \left[\int_{t_i}^{t_f} \dd{s}  \int_{t_i}^{s} \dd{u} \int_{t_i}^{u} \dd{v} 
\tilde{Q}(s) \tilde{Q}(u) Q(v) C(u,s,v) \right.\\
& \hspace{1.5in} \left. - Q(s) Q(u) \tilde{Q}(v) C(v,s,u) \right]
\end{split}
\end{equation}
A similar argument can be made for a term of order $N$.
There are 
$\displaystyle\frac{N!}{n! (N-n)!}$ ways to select $n$ indexes to be $+$, and $N-n$ indexes to be $-$.
By the time ordering the cumulant $G_N$ give the same if the times in the two groups are permuted within themselves
which can be done in $n!\times (N-n)!$ ways.
The contribution from $n$ forward paths ($Q$) and $m$ backward paths ($\tilde{Q}$) 
is thus $n$ time-ordered and $m$ reverse time-ordered integrals
multiplying the corresponding correlation function, which is 
\eqref{eq:cumulants-5} in main text.

Eq.~\eqref{eq:third-cumulant} can be analyzed further by considering in the mixed terms
the three ranges of $v$: less than $u$; between $u$ and $s$, and larger than $s$.
Renaming the variables so that times are always ordered $s>u>v$ this gives
\begin{equation}
\label{eq:third-cumulant-2}
\begin{split}
&\sum_{n+m=3}S^{(n,m)} = \\
& \qquad i \int_{t_i}^{t_f} \dd{s}  \int_{t_i}^{s} \dd{u} \int_{t_i}^{u} \dd{v}  
\left[Q(s) Q(u) Q(v) C(s,u,v) - Q(s) Q(u) \tilde{Q}(v) C(v,s,u) \right.  \\
& \qquad  - Q(s) \tilde{Q}(u) Q(v) C(u,s,v) - \tilde{Q}(s) Q(u) Q(v) C(s,u,v)+ \tilde{Q}(s) \tilde{Q}(u) Q(v) C(u,s,v) r \\
& \qquad \left. + \tilde{Q}(s) Q(u) \tilde{Q}(v) C(v,s,u) + Q(S) \tilde{Q}(u) \tilde{Q}(v) C(v,u,s) - \tilde{Q}(s) \tilde{Q}(u) \tilde{Q}(v) C(v,u,s) \right]
\end{split}
\end{equation}
Collecting terms with the same last entries one sees that this is
\begin{equation}
\label{eq:third-cumulant-3}
\begin{split}
&\sum_{n+m=3}S^{(n,m)} = i \int_{t_i}^{t_f} \dd{s} \left[Q(s)-\tilde{Q}(s)\right] \int_{t_i}^{s} \dd{u} \int_{t_i}^{u} \dd{v} \left[Q(u)Q(v)C(s,u,v) \right.\\
&\qquad \left. - Q(u) \tilde{Q}(v) C(v,s,u) - \tilde{Q}(u) Q(v) C(u,s,v) + \tilde{Q}(u) \tilde{Q}(v) C(v,u,s) \right]
\end{split}
\end{equation}
The third-order terms hence satisfy the general property of the Feynman-Vernon action
that if $Q(s)=\tilde{Q}(s)$ for all $s$ greater than $\tau$, then the action does not depend on
 $Q(s)$ or $\tilde{Q}(s)$ for $s>\tau$.
This conclusion also holds more generally: starting from \eqref{eq:cumulants-5} in main text
one can first insert $u_1$ in any of the intervals $[t_f;s_1]$,  $[s_1;s_2]$, \ldots, $[s_n;t_i]$,
then  $u_2$ in the same interval as $u_1$ or any one further down the list, and so on.
Each such insertion can be identified by a sequence $Z_{\tau_1},Z_{\tau_2},\ldots,Z_{\tau_N}$
where each symbol is $Q$ or $\tilde{Q}$, and the times are ordered $\tau_1>\tau_2>\cdots>\tau_N$. 
Consider now two cases that only differ by the symbol $Z_{\tau_1}$. The first case has $n$
symbols $Q$ and $N-n$ symbols $\tilde{Q}$ ($0<n\leq N$), and arises from inserting $(u_1,u_2,\ldots,u_{N-n})$ in $(s_1,s_2,\ldots,s_n)$
such that $u_1$ falls in one of the intervals $[s_1;s_2]$, \ldots, $[s_n;t_0]$.
The second case has on the other hand $n-1$
symbols $Q$ and $N-n+1$ symbols $\tilde{Q}$ ($0\leq n-1 < N$)
and arises from inserting $(u',u_1,u_2,\ldots,u_{N-n})$ in $(s_2,\ldots,s_n)$
such that $u'=s_1$ and $u_1,u_2,\ldots$ fall as in the first case.
The corresponding cumulant is in both cases $G_N(u_{N-n},\ldots,u_1,s_1,s_2,\ldots,s_n)$
which does not depend on the symbol of the largest time ($u'=s_1$).
Each such combination is therefore proportional to $Q(s_1)-\tilde{Q}(s_1)$.  

If written in terms of the $\zeta_{+}(s)=Q(s)+\tilde{Q}(s)$ and $\zeta_{-}(s)=Q(s)-\tilde{Q}(s)$,  Eq.  \eqref{eq:third-cumulant-3} can further be expressed as
\begin{equation}
\label{eq:third-cumulant-4}
\begin{split}
& \sum_{n+m=3}S^{(n,m)} = \frac{i}{4} \int_{t_i}^{t_f} \dd{s}  \zeta_{-}(s)  \\
& \qquad \times \int_{t_i}^{s} \dd{u} \int_{t_i}^{u} \dd{v}  \Big(\zeta_{+}(u) \zeta_{+}(v) A +\zeta_{+}(u)\zeta_{-}(v) B + \zeta_{-}(u) \zeta_{+}(v) C + \zeta_{-}(u) \zeta_{-}(v) D \Big)
\end{split}
\end{equation}
where the combined amplitudes can be written out as
\begin{align}
\label{eq:third-cumulant-5}
A &= \tr\Sb \left[ Y\Sb(s)\left(Y\Sb(u)Y\Sb(v)\rho - Y\Sb(u)\rho Y\Sb(v) - Y\Sb(v)\rho Y\Sb(u)+ \rho Y\Sb(u)Y\Sb(v)\right) \right] \\
B &= \tr\Sb \left[ Y\Sb(s)\left(Y\Sb(u)Y\Sb(v)\rho + Y\Sb(u)\rho Y\Sb(v) - Y\Sb(v)\rho Y\Sb(u)- \rho Y\Sb(u)Y\Sb(v)\right) \right] \\
C &= \tr\Sb \left[ Y\Sb(s)\left(Y\Sb(u)Y\Sb(v)\rho - Y\Sb(u)\rho Y\Sb(v) + Y\Sb(v)\rho Y\Sb(u)- \rho Y\Sb(u)Y\Sb(v)\right) \right] \\
D &= \tr\Sb \left[ Y\Sb(s)\left(Y\Sb(u)Y\Sb(v)\rho + Y\Sb(u)\rho Y\Sb(v) + Y\Sb(v)\rho Y\Sb(u)+ \rho Y\Sb(u)Y\Sb(v)\right) \right] 
\end{align}
By comparison, the standard Feynman-Vernon action can be written in a similar way as
\begin{equation}
\label{eq:second-cumulant}
\sum_{n+m=2}S^{(n,m)} = -\frac{1}{2} \int_{t_i}^{t_f} \dd{s}  \zeta_{-}(s)  \int_{t_i}^{s} \dd{u} \left(\zeta_{+}(u) A' +\zeta_{-}(u) B' \right)
\end{equation}
where 
\begin{align}
\label{eq:second-cumulant-2}
A' &= \tr\Sb \left[ Y\Sb(s)\left(Y\Sb(u)\rho - \rho Y\Sb(u)\right) \right]  \\
B' &= \tr\Sb \left[ Y\Sb(s)\left(Y\Sb(u)\rho + \rho Y\Sb(u)\right) \right] 
\end{align}
The contributions from the fourth order cumulants are analogously found to be 
\begin{equation}
\label{eq:fourth-cumulant}
\begin{split}
\sum_{n+m=4}S^{(n,m)} &= \int_{t_i}^{t_f} \dd{s} \left(Q(s)-\tilde{Q}(s)\right) 
\int_{t_i}^{s} \dd{u} \int_{t_i}^{u} \dd{v}  \int_{t_i}^{v} \dd{w} \Big(
Q(u) Q(v) Q(w) G_4(s,u,v,w) \\
&\quad
- \tilde{Q}(u) \tilde{Q}(v) \tilde{Q}(w) G_4(w,v,u,s)  - Q(u) Q(v) \tilde{Q}(w) G_4(w,s,u,v)  \\
&\quad 
- Q(u) \tilde{Q}(v) Q(w) G_4(v,s,u,w) -\tilde{Q}(u) Q(v) Q(w) G_4(u,s,v,w) \\
&\quad
+ \tilde{Q}(u) \tilde{Q}(v) Q(w) G_4(v,u,s,w) + \tilde{Q}(u) Q(v) \tilde{Q}(w) G_4(w,u,s,v) \\
&\quad +  Q(u) \tilde{Q}(v) \tilde{Q}(w) G_4(w,v,s,u) \Big)
\end{split}
\end{equation}
which can be re-written 
\begin{equation}
\label{eq:fourth-cumulant-2}
\begin{split}
\sum_{n+m=4}S^{(n,m)} &= \frac{1}{8} \int_{t_i}^{t_f} \dd{s} \zeta_{-}(s)  \\
& \times \int_{t_i}^{s} \dd{u} \int_{t_0}^{u} \dd{v} 
\int_{t_0}^{v} \dd{w}\sum_{pqr=\pm} \zeta_{p}(u) \zeta_{q}(v) \zeta_{r}(w) A_{pqr}(s,u,v,w,)
\end{split}
\end{equation}
and
\begin{align}
\label{eq:fourth-cumulant-3}
A_{+++} &= G_4(s,u,v,w) - G_4(w,v,u,s) - G_4(w,s,u,v) - G_4(v,s,u,w) \nonumber \\
& - G_4(u,s,v,w) + G_4(v,u,s,w) + G_4(w,u,s,v) + G_4(w,v,s,u) \nonumber\\ 
A_{++-} &= G_4(s,u,v,w) + G_4(w,v,u,s) + G_4(w,s,u,v) - G_4(v,s,u,w) \nonumber \\
& - G_4(u,s,v,w) + G_4(v,u,s,w) - G_4(w,u,s,v) - G_4(w,v,s,u) \nonumber\\ 
A_{+-+} &= G_4(s,u,v,w) + G_4(w,v,u,s) - G_4(w,s,u,v) + G_4(v,s,u,w) \nonumber \\
& - G_4(u,s,v,w) - G_4(v,u,s,w) + G_4(w,u,s,v) - G_4(w,v,s,u) \nonumber\\ 
A_{-++} &= G_4(s,u,v,w) + G_4(w,v,u,s) - G_4(w,s,u,v) - G_4(v,s,u,w) \nonumber\\
& + G_4(u,s,v,w) - G_4(v,u,s,w) - G_4(w,u,s,v) + G_4(w,v,s,u) \nonumber\\ 
A_{+--} &= G_4(s,u,v,w) - G_4(w,v,u,s) + G_4(w,s,u,v) + G_4(v,s,u,w) \nonumber\\
& - G_4(u,s,v,w) - G_4(v,u,s,w) - G_4(w,u,s,v) + G_4(w,v,s,u) \nonumber\\ 
A_{-+-} &= G_4(s,u,v,w) - G_4(w,v,u,s) + G_4(w,s,u,v) - G_4(v,s,u,w) \nonumber\\
& + G_4(u,s,v,w) - G_4(v,u,s,w) + G_4(w,u,s,v) - G_4(w,v,s,u) \nonumber\\ 
A_{--+} &= G_4(s,u,v,w) - G_4(w,v,u,s) - G_4(w,s,u,v) + G_4(v,s,u,w) \nonumber\\
& + G_4(u,s,v,w) + G_4(v,u,s,w) - G_4(w,u,s,v) - G_4(w,v,s,u) \nonumber\\ 
A_{---} &= G_4(s,u,v,w) + G_4(w,v,u,s) + G_4(w,s,u,v) + G_4(v,s,u,w) \nonumber\\
& + G_4(u,s,v,w) + G_4(v,u,s,w) + G_4(w,u,s,v) + G_4(w,v,s,u) \nonumber
\end{align}



\section{The path integral calculation of the generating function of heat for harmonic baths}
\label{app:path-integrals-harmonic-baths}
We restate from the main text of the paper that
the generating function of the energy change in one bath 
is defined as
\begin{equation}
\label{eq:FCS-APP}
G_{if}(\nu) = \Tr\Sb\matrixel{f}{e^{i\nu H\Sb} U e^{-i\nu H\Sb} \left(\dyad{i}\otimes \rho\Sb(\beta)\right)  U^{\dagger} }{ f}
\end{equation}
where $i$ and $f$ are the initial and final state of the system, 
$\rho_B(\beta)$ is the initial thermal state of the bath at inverse temperature $\beta$
and $\nu$ is the generating function parameter. 
In the Feynman-Vernon approach the two unitary operators
$U$ and $U^{\dagger}$, the 
final operator $e^{i\nu H_{B}}$, and the shifted initial thermal state of the 
$e^{-i\nu H_{B}}\rho_B(\beta)$
are all expressed as path integrals, and then the history of the bath is 
integrated out. 

For baths that are harmonic oscillators 
and for $\nu=0$ this was done exactly by Feynman and Vernon, giving 
\begin{equation}
P_{if}=G_{if}(\nu=0) = \int_{if} {\cal D}X {\cal D}Y e^{\frac{i}{\hbar}S\Ss[X]-\frac{i}{\hbar}S\Ss[Y]+\frac{i}{\hbar}S^{\nu=0}_i-\frac{1}{\hbar}S^{\nu=0}_r} 
\label{P-if}
\end{equation}
where  ${\cal D}X$ and $ {\cal D}Y$ are integrals over the forward and 
backward system paths,  
$S_i$ and $S_r$ are the two terms in the Feynman-Vernon action from 
integrating out the bath, and  $\int_{if}\left(\cdots\right)$ is a 
short-hand for projections on initial and final states. 
The Feynman-Vernon action is most commonly written as products of the sums and differences of the forward and backward paths,
$X+Y$ and $X-Y$. When $\nu$ can also be different from zero
it more convenient to instead write 
\begin{equation}
\label{eq:S_i+S_r}
\begin{split}
\frac{i}{\hbar}S^{\nu=0}_i[X,Y]-\frac{1}{\hbar} S_r^{\nu=0}[X,Y] &= 
   \frac{i}{\hbar}\int^t \int^s (XX'-YY') \kappa_i(s,s') \dd{s'} \dd{s}  \\
&-\frac{1}{\hbar} \int^t \int^s (XX'+YY') \kappa_r(s,s') \dd{s'} \dd{s}  \\
& +\frac{i}{\hbar}\int^t \int^t XY' \kappa_i(s,s') \dd{s'} \dd{s} \\
&+\frac{1}{\hbar} \int^t \int^t XY' \kappa_r(s,s') \dd{s'} \dd{s} 
\end{split}
\end{equation}
where primed (unprimed) quantities refer to time $s'$ ($s$) and the kernels $\kappa_i$ and $\kappa_r$ are given in \eqref{eq:kappa}
and \eqref{eq:k_r} in main text.

Now consider the generating function $G_{if}(\nu)$ of \eqref{eq:FCS-APP}.
Since the path integrals for this quantity are also all Gaussian
the path integrals pertaining to one harmonic oscillator reduce
to a four-dimensional integral 
\begin{equation}
\begin{split}
{\cal F}_b &= \int \dd{x_i} \dd{y_i} \dd{x_f} \dd{y_f} K^{f}(x_f,y_f,\hbar\nu) K(x_f,x_i,t_f-t_i;X)  \\
&\qquad \times K^*(y_f,y_i,t_f-t_i;Y) \left(K^{f}(x_i,y_i,-\hbar\nu+i\beta)\right)^*
\end{split}
\end{equation}
where $K^{f}$ is the free propagator of the bath and
$K(\cdot;X)$ is the propagator of the bath interacting linearly with an classical time-dependent field $X$,
and similarly for $K^*(\cdot;Y)$.
These propagators contain terms constant, linear and quadratic in the initial and final point
of each propagator. The quadratic terms are the same for the free and the interacting
propagators, the linear terms are integrals in the driving fields
($X$ and $Y$, respectively) and the constant term is 
one double integral
in $X$ minus 
one double integral
in $Y$.

By algebraic manipulation given in~\cite{Aurell2018} (appendix)
one can reduce ${\cal F}_b$
to $e^{\frac{i}{\hbar}S^{\nu=0}_i-\frac{1}{\hbar} S_r^{\nu=0}+{\cal J}^{(2)}+{\cal J}^{(3)}}$
where
\begin{align}
\label{eq:J2-def}
{\cal J}^{(2)} &= \frac{i}{2m\omega\hbar}\int^t\int^t \dd{s} \dd{s'} (XY'-X'Y)CC' \sin\omega(s-s') \left(\frac{yz'-y'z}{\Delta} -\frac{1}{2}\right) \\
\label{eq:J3-def}
{\cal J}^{(3)} &= \frac{i}{2m\omega\hbar}\int^t\int^t \dd{s} \dd{s'} (XY'+X'Y)CC' \cos\omega(s-s') \left(\frac{z'-y'}{\Delta} +\frac{i}{2}\coth\frac{\omega\hbar\beta}{2}\right)
\end{align}
The expressions in \eqref{eq:J2-def} and
\eqref{eq:J3-def}
depend on auxiliary parameters: 
$x=\cot(\omega t)$, $x'=\sin^{-1}(\omega t)$,
$y=\cot(\omega\hbar \nu)$, $y'=\sin^{-1}(\omega\hbar\nu)$,
$z=\cot(\omega\hbar(\nu-i\beta))$ and $z'=\sin^{-1}(\omega\hbar(\nu-i\beta))$.
$\Delta$ is the combination $2(z'y'-yz-1)$. 
Superscripts (2) and (3) in \eqref{eq:J2-def} and \eqref{eq:J3-def} 
are given for back-compatibility, and do not matter in the present discussion.
In~\cite{Aurell2018} the amplitude of ${\cal J}^{(2)}$ in \eqref{eq:J2-def} was incorrectly given
as $\left(\frac{y'z'-yz}{\Delta} -\frac{1}{2}\right)$;
the error was corrected in~\cite{Aurell2018-Erratum}.

We can now simplify to
\begin{align}
\Delta &= 2\sin^{-1}(\omega\hbar(\nu-i\beta))\sin^{-1}(\omega\hbar \nu)\left(1-\cosh(\omega\hbar\beta)\right) \\
\frac{yz'-y'z}{\Delta} &= \frac{1}{2} \cos(\omega\hbar \nu) + \frac{i}{2}\sin(\omega\hbar \nu)\coth(\frac{\omega\hbar\beta}{2})\\
\frac{z'-y'}{\Delta} &= -\frac{i}{2} \cos(\omega\hbar \nu) \coth(\frac{\omega\hbar\beta}{2}) + \frac{1}{2}\sin(\omega\hbar \nu)
\end{align}
and rewrite the integrands in
\eqref{eq:J2-def} and \eqref{eq:J3-def}.
 For terms proportional to $XY' CC'$ we have
\begin{equation}
\begin{split}
\hbox{Expr.} &= \sin\omega(s-s') \left[\frac{1}{2} \cos(\omega\hbar \nu) + \frac{i}{2}\sin(\omega\hbar \nu)\coth(\frac{\omega\hbar\beta}{2})\right] \\
&+ \cos\omega(s-s') \left[-\frac{i}{2} \cos(\omega\hbar \nu)\coth(\frac{\omega\hbar\beta}{2}) + \frac{1}{2}\sin(\omega\hbar \nu)\right]
\end{split}
\end{equation}
By trigonometry this is
$\frac{1}{2}\sin\omega(s-s'+\hbar\nu) 
-\frac{i}{2}\cos\omega(s-s'+\hbar\nu)\coth(\frac{\omega\hbar\beta}{2})$.
The terms proportional to  $X'Y CC'$ are similarly 
$-\frac{1}{2}\sin\omega(s-s'-\hbar\nu) 
-\frac{i}{2}\cos\omega(s-s'-\hbar\nu)\coth(\frac{\omega\hbar\beta}{2})$.
Exchanging labels and including the integrals and the prefactors in
\eqref{eq:J2-def} and \eqref{eq:J3-def} the cross-terms between the forward and backward paths for the generating function are hence

\begin{equation}
\label{eq:J-2-and-3-expr}
\begin{split}
\hbox{Cross-terms} &= \frac{i}{\hbar}\int^t \int^t XY' \sum_b C_b C'_b \frac{1}{2m_b\omega_b}\sin\omega_b(s-s'+\hbar\nu) \dd{s'}\dd{s} \\
              &+\frac{1}{\hbar} \int^t \int^t XY'  \sum_b C_b C'_b \frac{1}{2m_b\omega_b}\cos\omega_b(s-s'+\hbar\nu) \coth(\frac{\omega_b\hbar\beta}{2}) \dd{s'}\dd{s} 
\end{split}
\end{equation}
Comparing to the cross-terms in \eqref{eq:S_i+S_r} this is but a simple time shift of the arguments
of the sines and the cosines.

\section*{References}
\bibliographystyle{iopart-num}
\bibliography{refs}

\end{document}